\pdfoutput=1

\documentclass[11pt]{article}

\usepackage[]{acl}

\usepackage{times}
\usepackage{latexsym}

\usepackage[T1]{fontenc}
\usepackage{booktabs}
\usepackage{graphicx} 
\usepackage{hyperref}
\usepackage{color}
\usepackage{pythonhighlight}
\usepackage{graphicx}
\usepackage{pgfplots}
\usepackage[utf8]{inputenc}

\usepackage{microtype}
\usepackage{amsmath,lipsum}

\usepackage{inconsolata}
\usepackage{bigfoot}

\usepackage[english]{babel}
\usepackage{blindtext}

\title{eXplainable Bayesian Multi-Perspective Generative Retrieval}
\author{\textbf{EuiYul Song}$^{1\dagger}$~~~\textbf{Philhoon Oh}$^{2}$~~~\textbf{Sangryul Kim}$^{2}$~~~\textbf{James Thorne}$^{2}$ \smallskip \\ $^1${Samsung Electronics} , \texttt{euiyul.song@samsung.com} \\ $^2$KAIST AI , \texttt{\{philhoonoh,sangryul,thorne\}@kaist.ac.kr}}
\begin{document}
\maketitle
\begingroup\def\thefootnote{$\dagger$}\footnotetext{Work performed while at KAIST AI.}\endgroup
\begin{abstract}

Modern deterministic retrieval pipelines prioritize achieving state-of-the-art performance but often lack interpretability in decision-making. These models face challenges in assessing uncertainty, leading to overconfident predictions. To overcome these limitations, we integrate uncertainty calibration and interpretability into a retrieval pipeline. Specifically, we introduce Bayesian methodologies and multi-perspective retrieval to calibrate uncertainty within a retrieval pipeline. We incorporate techniques such as LIME and SHAP to analyze the behavior of a black-box reranker model. The importance scores derived from these explanation methodologies serve as supplementary relevance scores to enhance the base reranker model. We evaluate the resulting performance enhancements achieved through uncertainty calibration and interpretable reranking on Question Answering and Fact Checking tasks. Our methods demonstrate substantial performance improvements across three KILT datasets.
    
\end{abstract}
\section{Introduction}
With the integration of Transformers in information retrieval, there are significant advancements in knowledge-intensive language tasks \cite{petroni-etal-2021-kilt}. Nonetheless, the issue of bias in information retrieval poses challenges, encompassing information scarcity and overconfidence exhibited by models. These factors establish a glass ceiling in model accuracy, necessitating the use of extensive annotation and adversarial training. 

\begin{figure}[t]
    \centering
    \includegraphics[width=1.0\linewidth]{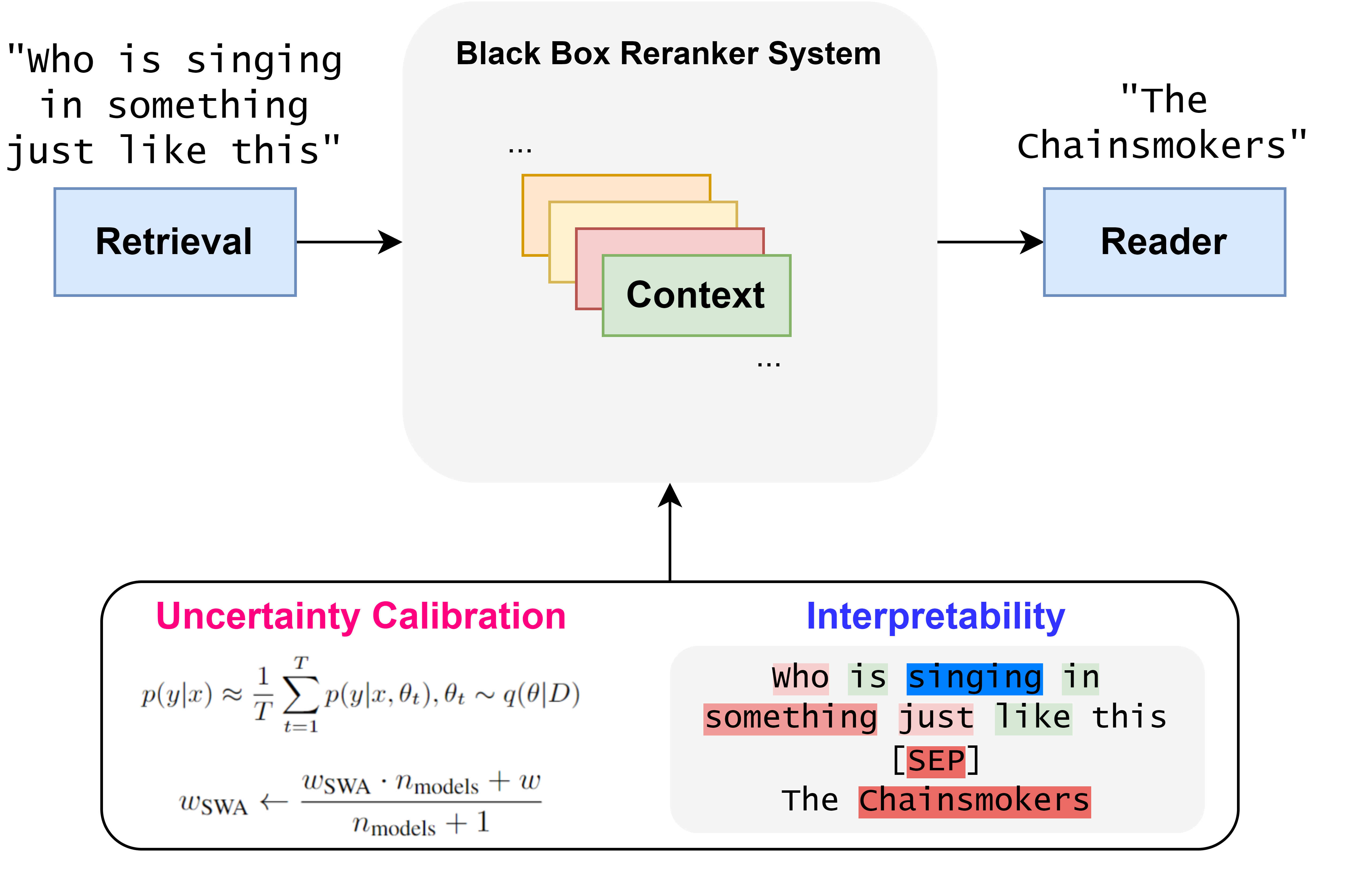}
    \caption{We add uncertainty calibration and explainability on the black box reranker system. We find that performance improves simply by applying two modules without a significant increase in inference latency.}
    \label{fig:abstract_arch}
\end{figure}

For instance, obtaining extensive, high-quality annotated data \cite{JMLR:v11:raykar10a, zaidan-callison-burch-2011-crowdsourcing, sabou-etal-2014-corpus} is crucial for debiasing and addressing overfitting concerns \cite{kaplan2020scaling}. However, the process of data annotation entails significant investments in time and resources for project management, Active Learning \cite{zhang2023survey}, data generation, data labeling, data visualization, and validation \cite{thorne-etal-2018-fever, schuster-etal-2019-towards, smit-etal-2020-combining, liu-etal-2022-wanli, wang2023far}.

Adversarial attacks and probing \cite{ribeiro-etal-2018-semantically, ribeiro-etal-2020-beyond, chen-etal-2020-seqvat, ivgi-berant-2021-achieving, qi-etal-2021-mind, miyato2021adversarial, perez-etal-2022-red, hartvigsen-etal-2022-toxigen, shi2023large} have been employed to enhance model generalization and robustness by introducing noise and addressing security concerns \cite{li-etal-2021-contextualized}. Nevertheless, adversarial training entails an increased computational cost, as it involves perturbation and carries the risk of overfitting to adversarial data.

To enhance the robustness of the existing retrieval pipeline while mitigating labeling and training costs, we introduce uncertainty calibration and explainability techniques. Our contributions are summarized as follows:
\begin{itemize}
\item  We implement multi-perspective retrieval, combining results when grounding an answer, resulting in a 2.91\% increase in downstream reader accuracy across three KILT datasets.

\item To address uncertainty in a reranker model, we employ Monte Carlo Dropout and Stochastic Weight Averaging. This calibration enhances reader accuracy by 0.77\% without incurring additional training costs. Reranking is performed using importance feature scores from LIME and SHAP, leading to a 1.73\% increase across two KILT datasets.

\item We conduct uncertainty-aware imputation pre-training of a reader using Stochastic Weight Averaging and Jensen Shannon Divergence. This approach yields a 0.68\% increase in downstream reader accuracy without introducing additional training and inference costs.
\end{itemize}
In summary, our approach encompasses reranking and reasoning using explainable feature scores, Bayesian deep learning, and multi-perspective retrieval. Together, these strategies significantly bolster the robustness of the retrieval pipeline. These advancements empower the retrieval pipeline to attain state-of-the-art performance, concurrently resulting in cost savings through reduced adversarial training time and decreased reliance on extensive data labeling.

\section{Related Work}

\subsection{Explainability and Interpretability}
With the increase in the complexity of neural network structures, it becomes difficult to discern the underlying rationale behind the model's outcomes. This phenomenon is often referred to as the "black box" problem \citep{hussain2019deep}, and ongoing research has been dedicated to enhancing the explainability and interpretability of the model's outcomes. 

One line of work to explain the model is to analyze the attention mechanism. Since attention calculates the distribution of the input, it can be viewed as an indicator \cite{kobayashi-etal-2020-attention} or at least as an attribute to identify the important feature. \cite{brunner2020identifiability}. However, the idea of using attention is criticized for its inconsistency, where attention distributions are misaligned even with the equivalent predictions \cite{jain2019attention}. \citealt{serrano2019attention} demonstrate that attention weights do not always correlate with the model prediction, and \citealt{liu2022rethinking} asserts that attention-based techniques may not accurately measure the level of impact contributed by each feature.


Another line of work for analyzing deep learning models emphasizes the interpretability aspect. Instead of directly analyzing the model itself, this approach involves training an auxiliary function, which is designed to provide explanations that offer insights into the behavior of the model. Therefore, the auxiliary function, often referred to as explanation models, should be simple enough for people to interpret the results \cite{ribeiro2016why, NIPS2017_8a20a862}. Also, features extracted by explainers can be served as supplementary inputs for models or leverage to enhance the performance of the model. \citealt{thorne2019generating} demonstrate that explanations generated by LIME \cite{ribeiro2016why} exhibit a higher level of similarity to human judgments compared to the attention thresholding approach in the Natural Language Inference task.

\subsection{Bayesian Deep Learning}
The field of Bayesian Deep Learning has witnessed extensive development, leveraging the probabilistic interpretation of deep learning models to estimate uncertainty. Variational Inference \citep{blundell2015weight, blei2017variational, Kingma_2019} has been extensively studied as a method to learn a probability distribution over the weights of a neural network. Concurrently, Stochastic Weight Averaging \citep{61aa9e9cc965421e82d7b7042c61abc8, gupta2020stochastic} involves averaging neural network weights sampled during stochastic gradient descent (SGD) training, enhancing optimization towards improved generalization. Monte Carlo Dropout \citep{10.5555/3045390.3045502} employs dropout in neural networks as a Bayesian approximation to calibrate uncertainty.

Snapshot Ensembling \citep{huang2017snapshot} undergoes multiple cycles of learning rate annealing during training, exploring and moving away from various local minima. It captures a snapshot at each minimum for ensemble use during test time. Fast Ensembling \citep{garipov2018loss} explores loss surfaces along trajectories of low loss using a cyclical learning rate schedule, subsequently averaging predictions from the traversed networks.

Deep Ensembles \citep{lakshminarayanan2017simple} entail training multiple deep neural networks with the same data and architectures but different initializations, followed by averaging their predictions to address model uncertainty. Gumbel Softmax \citep{jang2017categorical}, on the other hand, serves as a bridge between discrete and continuous spaces, providing a differentiable approximation for sampling from discrete distributions during training. The continuous evolution of these methods highlights the diversity within the Bayesian Deep Learning field, with each approach contributing unique value to uncertainty estimation.

\subsection{Retrieval Augmented Language Models}
Efficient document retrieval has long been dominated by statistical methods such as TF-IDF and BM25. With the rise of the Transformer era, new search methodologies like Dense Passage Retrieval (DPR) \citep{karpukhin-etal-2020-dense}, which leverages language model embeddings, as well as those employing Seq-to-Seq models \citep{lewis-etal-2020-bart, raffel2020exploring}, have emerged.

RAG \citep{lewis2020retrieval} integrates a pretrained retrieval with a pretrained Seq-to-Seq generator and demonstrates the feasibility of end-to-end fine-tuning. Building on this, Re2G \citep{glass-etal-2022-re2g} extends the work by adding a reranker, which merges contexts retrieved from ANN and BM25 indexes. Contriever \citep{izacard2022contriever} and Atlas \citep{izacard2022atlas} utilize contrastive learning, showcasing competitive results against existing models and displaying strong potential in unsupervised and multilingual retrieval tasks. 

In this work, we extensively leverage the work of \citealp{song2024re3val}, which utilizes reinforcement and reranking in a generative retrieval model. Its architecture builds upon previous works \citep{de2020autoregressive, thorne-2022-data}, providing a comprehensive and extended approach to interpreting remaining spaces in the field of generative retrieval.

\section{Method}
We present multi-perspective retrieval, consolidating results from both Re3val \cite{song2024re3val} and GENRE \cite{de2020autoregressive} when formulating an answer. Additionally, we explore the impact of Monte Carlo Dropout, Stochastic Weight Averaging, Deep Ensemble, and Snapshot Ensemble on Re3val's reranker. Introducing a novel reranking method, we utilize importance feature scores from LIME and SHAP as relevance scores. Furthermore, we undertake uncertainty-aware imputation pre-training for a reader, employing techniques such as Stochastic Weight Averaging and Jensen Shannon Divergence.




\subsection{Bayesian Context Reranker}
Re3val utilizes a deterministic binary classification approach in its context reranker. However, their analysis does not consider the impact of calibrating model uncertainty. To overcome this limitation, we explore various Bayesian Deep Learning techniques: Deep Ensemble \citep{lakshminarayanan2017simple}, Snapshot Ensemble \citep{huang2017snapshot}, Stochastic Weight Averaging \citep{61aa9e9cc965421e82d7b7042c61abc8}, and Monte Carlo Dropout \citep{10.5555/3045390.3045502}. Our objective is to assess the effects of these uncertainty calibration techniques on the performance of the context reranker.

\subsubsection{Deep Ensemble}
Let $x \in N^K$ represent the input token, and $y \in \{0, 1\}$ represent the corresponding binary label, where $K$ denotes the vocabulary size. We consider an ensemble of context rerankers with $M$ rerankers in the ensemble. The ensemble comprises parameters $\theta_m$ for $m=1$ to $M$, and $p_\theta(y|x)$ represents the probabilistic distribution over labels.

During the training phase, we initialize the parameters $\theta_1, \theta_2, ..., \theta_M$ with different random seeds. At test time, we combine the predictions by averaging their probabilistic distributions over labels as follows:
\begin{equation} \small p(y|x) = \frac{1}{M} \sum_{m=1}^{M} p_{\theta_m}(y|x, \theta_m) \end{equation}

\subsubsection{Snapshot Ensemble}
In the Snapshot Ensemble technique, we incorporate a cyclic annealing scheduler to adjust the learning rate during training. Let $T$ denote the total number of training iterations, $\alpha_0$ represent the initial learning rate, $\theta$ denote a parameter, and $M$ signify a cycle. The cyclic annealing scheduler is defined as follows:
\begin{equation}
\label{equation:snapshot}
\small \alpha(t) = \frac{\alpha_0}{2}\left(\cos\left(\frac{\pi \text{ mod}(t-1, \lceil T/M \rceil)}{\lceil T/M \rceil}\right) + 1\right) \end{equation}
During testing, we average the prediction probabilities of the last 3 out of $M$ cycles.
\subsubsection{Stochastic Weight Averaging}

To improve the stability of Re3val's context reranker, we have implemented Stochastic Weight Averaging (SWA) on the model weights using the Adam optimizer. The weight-averaging formula is as follows:

\begin{equation} w_{\text{SWA}} \leftarrow \frac{w_{\text{SWA}} \cdot n_{\text{models}} + w}{n_{\text{models}} + 1} \end{equation}

Here, $w_{\text{SWA}}$ represents the averaged weights, $n_{\text{models}}$ is the number of models included in the averaging, and $w$ represents the weights of the current model being added to the average.

\subsubsection{Monte Carlo Dropout}
We utilize Monte Carlo Dropout, which randomly deactivates neurons in a neural network during training and inference. This approach differs from conventional dropout, which does not deactivate neurons during the inference stage. By applying dropout during inference, Monte Carlo Dropout samples from an approximate posterior distribution denoted as $q(\theta | D)$, where $D$ represents the training examples. Using Monte Carlo Dropout, our prediction probabilities are modified accordingly.

\begin{equation}
\small
    p(y|x) \approx \frac{1}{T} \sum_{t=1}^{T} p(y | x, \theta_t), \theta_t \sim q(\theta | D)
\end{equation}
\begin{figure*}[t]
    \centering
    \includegraphics[width=0.9\linewidth]{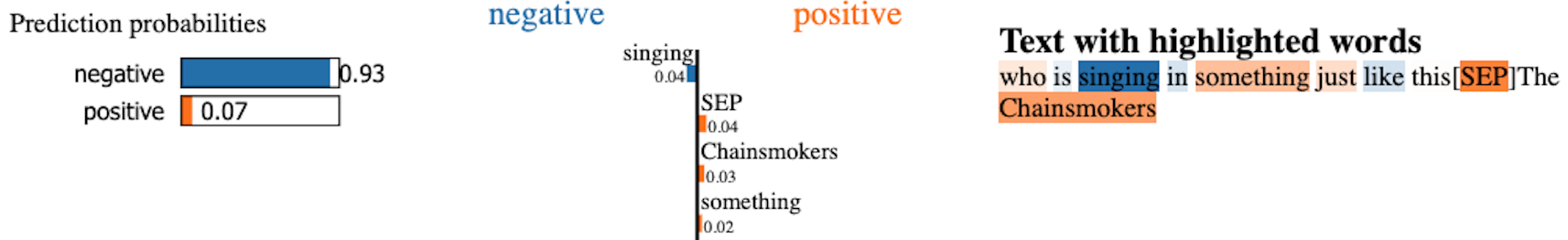}
    \caption{LIME visualization of the binary context reranker. Among the analyzed features, \textbf{Chainsmokers} appears to be the second most positive feature following SEP.}
    \label{fig:lime}
\end{figure*}

\begin{figure*}[t]
    \centering
    \includegraphics[width=0.9\linewidth]{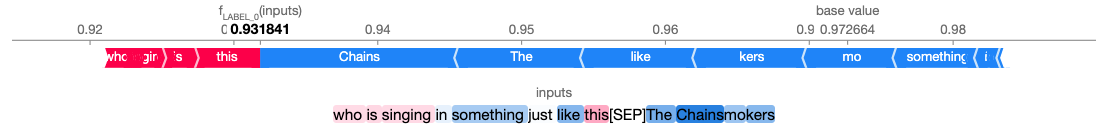}
    \caption{SHAP visualization for the binary context reranker reveals that the feature \textbf{Chains} has the most significant impact on determining the context as relevant.}
    \label{fig:shap}
\end{figure*}

\begin{figure}[t]
    \centering
    \includegraphics[width=0.9\linewidth]{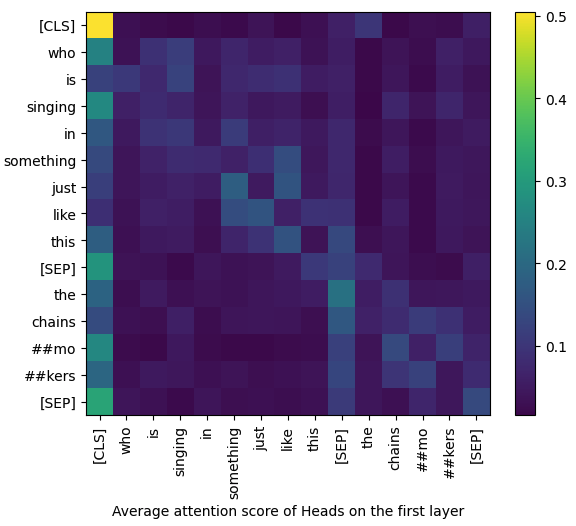}
    \caption{Average attention score on the first layer.}
    \label{fig:attention-first}
\end{figure}
\begin{figure}[t]
    \centering
    \includegraphics[width=0.9\linewidth]{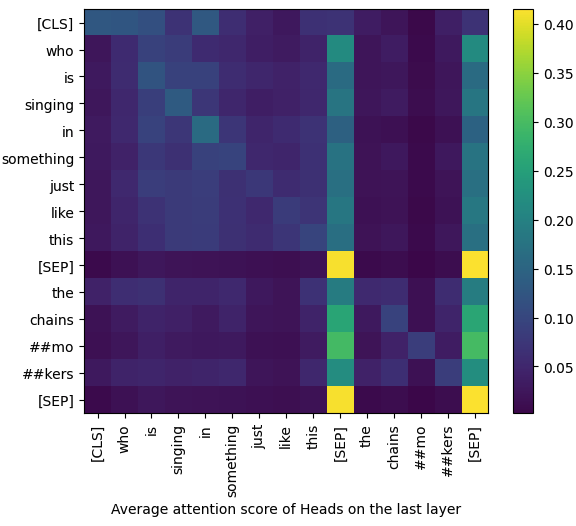}
    \caption{Averaged attention score on the last layer.}
    \label{fig:attention-last}
\end{figure}
\subsection{eXplainable Context Reranker}
Reranking candidate passages plays a pivotal role in elevating the performance of knowledge-intensive downstream tasks. Ensuring the relevance of contexts to queries is crucial for achieving enhanced results. To assess the effectiveness of Re3val's Context Reranker, we conducted an analysis of the visual representation of the context using LIME \cite{ribeiro2016why}, SHAP \cite{NIPS2017_8a20a862}, and attention values. In Figure~\ref{fig:lime}, LIME results indicate that \underline{Chainsmokers} is the second most influential factor for the positive label given the query "who is singing in something just like this." Similar findings are observed with SHAP, as illustrated in Figure~\ref{fig:shap}, where \underline{Chains} emerges as the most influential factor for the prediction.

Contrary to these findings, attention scores in Figure~\ref{fig:attention-first} and Figure~\ref{fig:attention-last} do not provide information related to the prediction label. Each figure represents the average score of attention heads on the first and last layer, respectively. Individual attention scores are included in the Appendix \ref{sec:appendix:attention}.

Following the analysis, we leverage features extracted from LIME and SHAP to assess their potential in enhancing downstream task performance. For LIME, features are extracted using the top-5 contexts, and the top-10 passages are subsequently reranked based on these features. Conversely, for SHAP, features are extracted from the top-1 passage due to its resource-intensive nature, and the top-10 passages are re-ranked accordingly.


\subsection{Uncertainty Aware Fusion in Decoder}
\subsubsection{Pre-training}
\label{method:uncertainty:pre}
Re3val incorporates imputation pre-training (Re3val$_I$) as a solution to the sparsity challenge posed by Missing Not At Random (MNAR) gold data. Nevertheless, exploring the impact of introducing stochasticity to reduce uncertainty in Re3val$_I$ requires further investigation.

Our initial approach involves enhancing Re3val$_I$ through Stochastic Weight Averaging (SWA) while intentionally introducing asymmetry in the number of particles between gold and imputed contexts. Subsequently, we aim to examine the effects of particle symmetry during imputation pre-training by ensuring an equal distribution of imputed and gold data samples in the SWA process.

Furthermore, minimizing the dissimilarity between the imputed and gold data distributions in our training objective can further reduce entropy. Let $f_\theta$ represent FiD \cite{izacard2020leveraging}, $x_{i}$ denote imputed contexts, $z_{i}$ signify gold contexts, $M$ represent the vocabulary size, $g_\phi$ represent the FiD encoder, $P = \frac{e^{g_\phi(x_{i})}}{e^{g_\phi(x_{i})} + e^{g_\phi(z_{i})}}$, and $Q = \frac{e^{g_\phi(z_{i})}}{e^{g_\phi(x_{i})} + e^{g_\phi(z_{i})}}$. Our training objective for pre-training becomes:
\begin{equation} \small{ L(\theta) = \frac{1}{M}\sum_{i=1}^{M} y_{i} \log f_\theta(x_{i}, z_{i}) + JSD(P||Q)} \end{equation}
Here, JSD (Jensen-Shannon Divergence) \citep{menendez1997the} is defined as:
\begin{equation}
\label{equation:jensen}
\small
JSD = H(\frac{1}{2}(P+Q)) - \frac{1}{2}(H(P) +H(Q)) 
\end{equation}
In the above equation, $H(X) = -\sum_{x \in X}{xlog(x)}$.

\subsubsection{Multi-Perspective Retrieval}
\label{method:uncertainty:multi}
Re3val offers empirical support for the idea that mutual information at the retrieval level diminishes entropies and aids in the reranking process. However, the influence of mutual information at the reader level necessitates additional exploration. Additionally, the inclusion of correctly added passages containing detrimental information leads to alterations in the reader model's predictions \cite{oh-thorne-2023-detrimental}. In an effort to investigate this phenomenon, we retrieve page titles from two indexes, namely GENRE and Re3val, to assess the impact of mutual information during reasoning. Mathematically, we express the impact as follows:

\begin{equation}
\small
I(X;Y) = \sum_{x \in \mathcal{X}} \sum_{y \in \mathcal{Y}} P(x, y) \log \frac{P(x, y)}{P(x)P(y)}
\end{equation}

In this equation, $X$ represents the top-2 reranked GENRE contexts, $Y$ represents the top-3 Re3val contexts. By examining the combined likelihood of GENRE and Re3val contexts, the mutual information $I(X;Y)$ enables us to understand the interdependence of these variables. Consequently, our reader utilizes this mutual information to minimize uncertainty in Re3val and GENRE contexts, thereby enhancing the overall reader performance.

Additionally, we generate a contrastive version of each query to evaluate whether the contexts retrieved using the contrastive query can reduce the entropies of the contexts obtained with the original query while reading.

\section{Experiment}

\subsection{Data}

To facilitate the training of the Bayesian Context Reranker and the execution of FiD pre-training, we utilize a dataset comprising 48,000 instances sourced from Natural Questions (Kwiatkowski et al., 2019), HotpotQA (Yang et al., 2018), and FEVER (Thorne et al., 2018), which is part of the KILT dataset (Petroni et al., 2021). These instances are uniformly sampled for training purposes. Additionally, for asymmetric SWA pre-training in conjunction with FiD, we integrate an additional 48,800 instances uniformly sampled from WoW (Dinan et al., 2019) and TriviaQA (Joshi et al., 2017). A detailed summary of the training and development data quantities is in Table \ref{tab:data}.

\subsubsection{Contrastive Query Generation} For contrastive query generation explained in Section 3.4.2, we use OpenAI gpt-3.5-turbo \cite{brown2020language} to generate a new query with the prefix "Generate a contrastive version of the following {query}. Output: ". For instance, the query "When did the new Maze Runner movie come out?" augments to "When is the release date of the old Maze Runner movie?".

\subsubsection{Contexts}
We use GENRE \citep{de2020autoregressive} pre-trained\footnote{\href{GENRE}{https://huggingface.co/facebook/genre-kilt}} on 11 KILT \citep{petroni-etal-2021-kilt} datasets to retrieve the top 5 page titles with a query and a contrastive query. Furthermore, we utilize Re3val to retrieve other top-5 page titles with a query. Following the retrieval process, we individually input the page titles from each index into the Re3val Context Reranker to obtain the top 5 contexts related to a query.

\subsection{Hyperparameters}
The standard hyperparameter configurations and hardware settings utilized for all tasks are delineated in Table \ref{tab:hyperparameter}.
\subsection{Metrics}
Evaluation metrics for downstream reading tasks vary based on the specific task at hand. For instance, question-answering tasks are assessed using metrics like Exact Match (EM) and F1 scores, while dialogue tasks utilize measures such as ROUGE-L and F1 scores. Moreover, fact-checking tasks are evaluated using the Accuracy (AC) score. 

\subsection{Bayesian Context Reranker}
We employ three samples for averaging probabilities in Deep Ensemble, Snapshot Ensemble, and Monte Carlo Dropout. In the case of stochastic weight averaging, we average the weights over Adam's entire trajectory on a cross-encoder \cite{devlin2019bert}. The outcomes of all these variations are then fed into the Re3val$_I$, trained on T5-large \cite{raffel2020exploring}. This model is trained with five contexts, including gold contexts and imputed DPR contexts from the DPR multi-set\footnote{\href{DPR}{https://dl.fbaipublicfiles.com/dpr/checkpoint/retriver/multiset}}.
\subsection{eXplainable Context Reranker}
To evaluate the effectiveness of the explainers as feature extractors, we conduct experiments using LIME\footnote{\href{LIME}{https://github.com/marcotcr/lime}} and SHAP\footnote{\href{LIME}{https://github.com/slundberg/shap}} implemented by authors. As these modules involve sampling, they are resource-intensive. To address this issue, we limit the number of contexts used for feature extraction to top-5 for LIME and top-1 for SHAP. Then we construct a feature score dictionary, where tokens serve as keys, and the values are corresponding scores calculated by each module. Lastly, the top 10 contexts are sorted by aggregating the scores of features presented in each context. For LIME, we extract 10 features using 150 samplings while we employ the default settings for SHAP.
\subsection{Uncertainty Aware Fusion in Decoder}
\subsubsection{Pre-training}
To augment Re3val's Fusion in Decoder pre-training through the introduction of stochasticity, we have employed a symmetrical technique for integrating both gold and DPR contexts. Our methodology entails initially inputting three gold contexts, succeeded by three DPR contexts. In cases where an instance possesses fewer than three gold contexts, we exclusively provide a query without a context for the remaining N encoders, with N calculated as the difference between 3 and the available number of gold contexts for that specific instance. Subsequently, we compute the average weights from checkpoints over the last 30\% of the training budget.

\subsubsection{Multi-Perspective Retrieval}
We assess the impact on performance by combining the contexts from GENRE with a query and contrastive query and the contexts from Re3val with a query. In total, we consider reading five contexts. During inference, we use the pre-trained FiD model that was trained using five contexts, including the gold contexts and imputed DPR contexts for a query with less than five gold contexts as Re3val \citep{song2024re3val}. All the contexts above are concatenated with the query and the page title.

\section{Result}
\begin{table*}[h]
\centering
\label{tab:final_score2}
{\small
\begin{tabular}{c|ccccc|c}
\toprule
& \multicolumn{4}{c}{QA} & \multicolumn{1}{c}{FC} \\
\textbf{Dataset} & \multicolumn{2}{c}{\textbf{NQ}} & \multicolumn{2}{c}{\textbf{HoPo}} & \multicolumn{1}{c}{\textbf{FEV}} & \\
\toprule
\textbf{Model} & \textbf{EM} & \textbf{F-1} & \textbf{EM} & \textbf{F-1} & \textbf{AC} & \textbf{AVG} \\
\midrule
Re3val$_I$ & 34.58 & 44.02 & 32.25 & 42.56 & 78.00 & 48.28 \\
\midrule
\textbf{+Lime} & 35.21 & 45.09 & 28.25 & 38.43 & \underline{79.10} & 47.52 \\
\textbf{+Shapley} & \textbf{36.38} & \textbf{46.32} & 26.30 & 35.93 & \textbf{79.66} & 47.45 \\
\midrule

\textbf{+Deep Ensemble} & 34.90 & 44.87 & \underline{32.37} & \underline{42.76} & 78.29 & 48.52 \\
\textbf{+Snapshot Ensemble} & 33.87 & 43.82 & 32.21 & 42.64 & 78.70 & 48.26\\
\textbf{+Stochastic Weight Averaging} & 35.04 & 44.67 & \textbf{32.48} & \textbf{42.94} & 78.42 & \underline{48.64} \\
\textbf{+Monte Carlo Dropout} & \underline{36.34} & \underline{45.92} & 32.32 & 42.60 & 78.49 & \textbf{49.05}\\
\bottomrule
\end{tabular}
}
\caption{Effect of Bayesian and eXplainable Context Reranking on the downstream reading task. Evaluation is conducted on the pre-trained Re3val$_{I}$ without any fine-tuning. LIME, SHAP, Deep Ensemble, Snapshot Ensemble, Stochastic Weight Averaging, and Monte Carlo Dropout are applied to the Re3val context reranker.}
\label{tab:table-2}
\end{table*}
\begin{table*}[h]
\centering
\label{tab:swa_assymetry}
{\small
\begin{tabular}{c|ccccccccc|c}
\toprule
  & \multicolumn{6}{c}{QA} & \multicolumn{1}{c}{FC} & \multicolumn{2}{c}{Dial} \\
\textbf{Dataset} & \multicolumn{2}{c}{\textbf{NQ}} & \multicolumn{2}{c}{\textbf{HoPo}} & \multicolumn{2}{c}{\textbf{TQA}}& \multicolumn{1}{c}{\textbf{FEV}} & \multicolumn{2}{c}{\textbf{WoW}}&\\
\midrule
\textbf{Model} & \textbf{EM} & \textbf{F-1} & \textbf{EM} & \textbf{F-1} & \textbf{EM} & \textbf{F-1} & \textbf{AC} & \textbf{RL} & \textbf{F-1} & \textbf{AVG}\\
\midrule
Re3val$_I$ & \textbf{60.10} & \textbf{70.70} & \textbf{51.20} & \textbf{64.05}  & \textbf{60.60} & \textbf{75.43} & 94.50 & 24.79 & 24.76 & 58.24\\
\midrule
\textbf{+SWA} & 58.60 & 69.71 & 39.40 & 57.18 & 59.90 & 74.77 & \textbf{95.10} & \textbf{25.31} & \textbf{25.26} & 55.66 \\
\bottomrule
\end{tabular}
}
\caption{The development set outcomes for Re3val$_I$ pre-training with SWA with sample asymmetry. The evaluation involved 5 contexts, comprising gold and imputed DPR contexts for queries with fewer than 5 gold contexts. Notably, the performance of SWA degrades when there is an asymmetry in the count between gold and DPR contexts.}
\label{tab:swa_assymetry}
\end{table*}
\begin{table*}[h]
\centering
\label{tab:swa_symmetry}
{\small
\begin{tabular}{c|ccccc|c}
\toprule
  & \multicolumn{4}{c}{QA} & \multicolumn{1}{c}{FC} \\
\textbf{Dataset} & \multicolumn{2}{c}{\textbf{NQ}} & \multicolumn{2}{c}{\textbf{HoPo}} & \multicolumn{1}{c}{\textbf{FEV}} &\\
\midrule
\textbf{Model} & \textbf{EM} & \textbf{F-1} & \textbf{EM} & \textbf{F-1} & \textbf{AC} & \textbf{AVG}\\
\midrule
Re3val$_I$ & 60.32 & 71.4 & 58.65 & 73.22  & 92.05 & 70.34\\
\midrule
\textbf{+JSD} & 60.47 & 71.52 & 59.03 & 73.48 & 92.68 &  70.73\\
\textbf{+SWA} & \textbf{60.75} & \underline{72.00} & \underline{59.30} & \underline{73.57} & \underline{92.73} & \underline{70.93} \\
\textbf{+JSD+SWA} & \underline{60.72} & \textbf{72.11} & \textbf{59.53} & \textbf{73.75} & \textbf{92.82} & \textbf{71.02} \\
\bottomrule
\end{tabular}
}
\caption{Development set results for our Uncertainty Aware FiD Pre-training Strategy. Evaluated with 3 gold and 3 imputed contexts per query, using the same pre-processing strategy as in Section \ref{method:uncertainty:pre}}
\label{tab:swa_symmetry}
\end{table*}
\begin{table*}[h]
\centering
\label{tab:final_score}
{\small
\begin{tabular}{cc|ccccc|c}
\toprule
  & & \multicolumn{4}{c}{QA} & \multicolumn{1}{c}{FC} \\
\textbf{Dataset} & & \multicolumn{2}{c}{\textbf{NQ}} & \multicolumn{2}{c}{\textbf{HoPo}} & \multicolumn{1}{c}{\textbf{FEV}} &\\
\toprule
\textbf{Model} & \textbf{|C|} & \textbf{EM} & \textbf{F-1} & \textbf{EM} & \textbf{F-1} & \textbf{AC} & \textbf{AVG} \\
\midrule
Re3val & 5& 34.58 & 44.02 & 32.25 & 42.56 & 78.00 & 48.28\\
\midrule
GENRE & 5 &31.05 & 40.00 &  24.96 & 34.17  & 78.60 & 44.87\\
$\neg$GENRE & 5 & 31.27 & 39.91 & 24.32 & 33.27 & 79.49 & 45.03 \\

GENRE$;\neg$GENRE & 3;2& 35.25 & 44.81 & 26.73 & 36.38 & 79.59 & 46.19 \\
\midrule
Re3val$;$GENRE &3;2& \underline{39.23} & \textbf{49.37} & 34.02 & \textbf{44.78} & \textbf{80.03} & \underline{51.09} \\
Re3val$;$GENRE$;\neg$GENRE &2;2;1& \textbf{39.62} & \underline{49.31} & \textbf{34.11} & 44.77 & \underline{79.85} & \textbf{51.19}\\
\bottomrule
\end{tabular}
}
\caption{Development set results on a Reader with contexts retrieved from different indices. \textbf{|C|} indicates the number of contexts used for reading, with a separator "$;$" separating the number of contexts from the corresponding index. "$\neg$" indicates the utilization of contexts retrieved with a contrastive query.
}
\label{tab:final_score}

\end{table*}
\subsection{Bayesian Context Reranker}
Due to time and resource limitations, we were unable to incorporate uncertainty metrics such as negative log-likelihood, expected calibration error \citep{guo2017on}, and Brier score \citep{lakshminarayanan2017simple} into our evaluation of Bayesian Context Reranker variants. Nevertheless, our results in Table \ref{tab:table-2} indicate an increase in reader performance when utilizing retrieved contexts from all our Bayesian Context Reranker variants, except for the one using Snapshot Ensemble. This finding suggests that our uncertainty calibration methodologies, excluding the Snapshot Ensemble, effectively enhance the reranking function. It is worth noting that all the variants, except Deep Ensemble, do not require additional training time and memory. Thus, these methodologies are promising for advancing the next generation of retrieval pipelines.

\subsection{Explainable Feature Reranker}
The downstream performances for using LIME and SHAP for rearnking the contexts are described in Table \ref{tab:table-2}. Surprisingly, SHAP achieves the highest score in the Natural Question and FEVER dataset, followed by Monte Carlo Dropout and LIME. These results suggest that explainer models are effective in identifying relevant features existing in the contexts. However, in the case of HotPotQA, neither LIME nor SHAP performs as well as other approaches. We speculate that this discrepancy is due to the multi-hop nature of HotPotQA, where relevant information is not confined to a single passage. Consequently, extracting features independently from a single context may fail to capture crucial features in such scenarios.

\subsection{Uncertainty Aware Fusion in Decoder}
\subsubsection{Pre-training}
Stochastic Weight Averaging with asymmetry between gold and imputed samples does not yield the desired performance improvements according to Table \ref{tab:swa_assymetry}. We hypothesize that the uneven number of particles between the imputed and gold data introduces bias during the model averaging.

Specifically, when there are more imputed data samples than gold data samples during the reading phase, the imputed data can disproportionately influence the final weights through model averaging. This imbalance can negatively impact performance since the imputed data may not be relevant to the query. We ensure equal imputed and gold data samples during SWA to mitigate this issue.

Table \ref{tab:swa_symmetry} presents compelling evidence of consistent performance improvement when employing Stochastic Weight Averaging on Re3val$_{I}$ with an equal number of particles derived from two distinct data distributions. This equitable allocation of particles ensures a balanced influence from both data sources on the final model averaging weights, effectively mitigating potential biases in imputed contexts.

In addition, incorporating Jensen-Shannon Divergence (JSD) between probability distributions of two encoded contexts in the Fusion in Decoder facilitates capturing similarity and distribution alignment. Our loss function (\ref{equation:jensen}) promotes more confident predictions when reading contexts from the two distributions. Significantly, as evidenced by the results in Table 3, the inclusion of JSD leads to improved reasoning capabilities, further highlighting the effectiveness of Jensen-Shannon Divergence in boosting confidence when processing samples from different distributions.

\subsubsection{Multi-Perspective Retrieval}
The evidence presented in Table \ref{tab:final_score} strongly supports a substantial improvement in downstream reader performance by integrating information from Re3val and GENRE contexts. This enhancement demonstrates the mutually complementary nature of the information obtained from both indexes, effectively reducing uncertainty. Additionally, combining contexts retrieved using both a contrastive query and a regular query leads to reduced entropies associated with their respective contexts, as evidenced by the performance results. These findings also highlight the effectiveness of Re3val's context reranker in accurately reranking the relevant contexts at the top.

\section{Conclusion}

Our study emphasizes the crucial role of uncertainty calibration and interpretability in strengthening the robustness of information retrieval pipelines. By leveraging Bayesian methodologies such as Stochastic Weight Averaging and Monte Carlo Dropout, we achieved a notable improvement in performance for a deterministic cross-encoder reranker across three KILT datasets without incurring additional training time. Furthermore, incorporating LIME and SHAP as relevance measures for the cross-encoder reranker resulted in a significant performance boost across the two datasets. Our implementation of the Stochastic FiD pre-training strategy and multi-perspective retrieval surpasses the performance of the existing base model but also demonstrates our pioneering use of uncertainty calibration when grounding an answer. This innovative approach contributes to a deeper understanding of uncertainty calibration and interpretability, promoting the development of more robust information retrieval systems.

\section*{Limitations}
While our contribution is sound and clear, there exists a challenge in incorporating explainable modules like LIME and SHAP. As \citealt{thorne2019generating} points out, the computational cost of generating explanations for black-box models is considerably high, making it impractical for test time inference. Additionally, in the multi-hop setting, these techniques demonstrate lower performance compared to other methods. Therefore, further research is required to explore the integration of black-box explainers into multi-hop environments.
\section*{Ethics Statement}
We make use of datasets sourced from Natural Questions, TriviaQA, HotpotQA, FEVER, and Wizard of Wikipedia. These datasets play a crucial role in the KILT benchmark and are derived from the KILT knowledge source, based on the English Wikipedia dump on August 1st, 2019. It's important to note that these datasets might include instances of inaccurate or misunderstood information, leading to the potential generation of biased, toxic, or fabricated content. Furthermore, the employment of language models like T5 and BERT during training introduces ethical risks embedded within the internal parameters of these models. As a result, researchers should exercise caution when utilizing our paper and its outputs, establishing appropriate policies to address potential ethical risks that may arise when deploying these models in real-world production scenarios.
\section*{Acknowledgement}
We appreciate ChatGPT 3.5's assistance in correcting writing errors.


\bibliography{output}

\onecolumn
\appendix

\section{Appendix}
\subsection{Hyperparameters}
\begin{table*}[h]
\centering
{\small
\begin{tabular}{cccccc}
\toprule
\textbf{Configuration} & \textbf{Context Reranker} & \textbf{FiD}  \\
\midrule
parameters & 110M & 770M \\
base model & bert-base & t5-large \\
runs & single & single \\
learning rate  &5e-5 & 1e-4 \\
scheduler &  linear & constant \\
warmup ratio  & 0 & 0 \\
eval steps ratio & 10\% & 10\% \\
batch size  &  1200* & 32*  \\
max seq length & 250* & 250* \\
max target length & 50 & 50 \\
epoch & 4 & 5*  \\
train beam size  & 1 & 1  \\
eval beam size  & 1 & 1  \\
dropout rate  &  0 & 0  \\
optimizer & AdamW& AdamW \\
gpu & A100 & A100 \\
early stopping steps & 4 & 4  \\

\bottomrule
\end{tabular}
}
\caption{The hyperparameter and hardware configurations employed in our study have been detailed earlier. Instances marked with asterisks (*) signify variations specifically applied to the WoW dataset. For WoW, we trained with a maximum sequence length twice that of the max seq length and half the batch size above.}
\label{tab:hyperparameter}
\end{table*}
\subsection{Data}
\begin{table*}[h]
\centering
{\small
\begin{tabular}{cccccc}
\toprule
\textbf{Splits} & \textbf{NQ} & \textbf{TriviaQA} & \textbf{HotpotQA} & \textbf{FEVER} & \textbf{WoW}  \\
\midrule
train & 48,000 & 48,000 & 48,000 & 48,000 & 48,000 \\
dev & 2,837 & 5,359 & 5,600 & 10,444 & 3,054 \\

\bottomrule
\end{tabular}
}
\caption{The number of training and development data.}
\label{tab:data}
\end{table*}
\subsection{Attention Score}
\label{sec:appendix:attention}
\begin{figure*}
    \centering
    \includegraphics[width=1\linewidth]{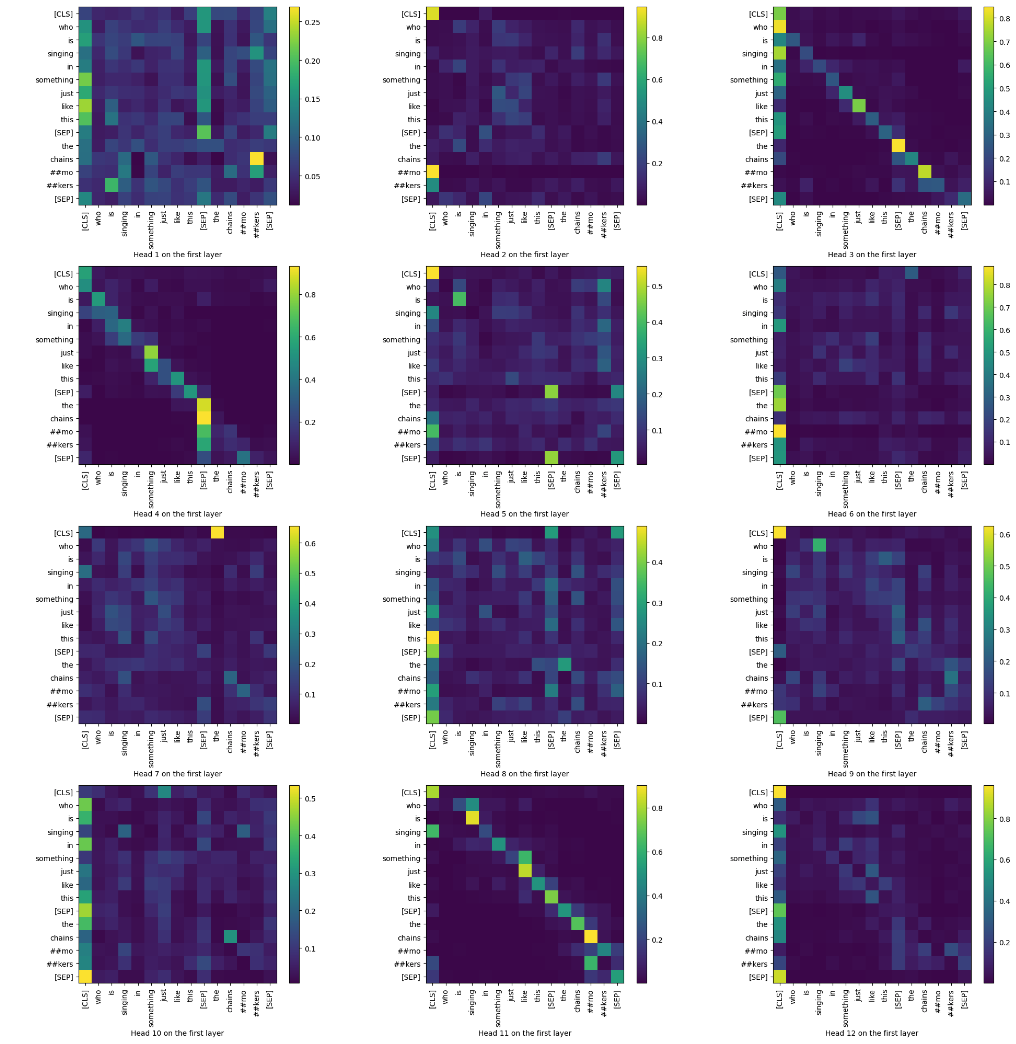}
    \caption{The attention scores of the heads in the initial layer of the binary context reranker. Tokens are self-attended on Heads 3, 4, and 11. Heads 1, 2, 6, 8, 9, 10, and 12 primarily focus on the [CLS] token. However, no discernible pattern is observed for the remaining heads. This does not offer meaningful information regarding the context.}
    \label{fig:asfist}
\end{figure*}
\pagebreak
\begin{figure*}
    \centering
    \includegraphics[width=1\linewidth]{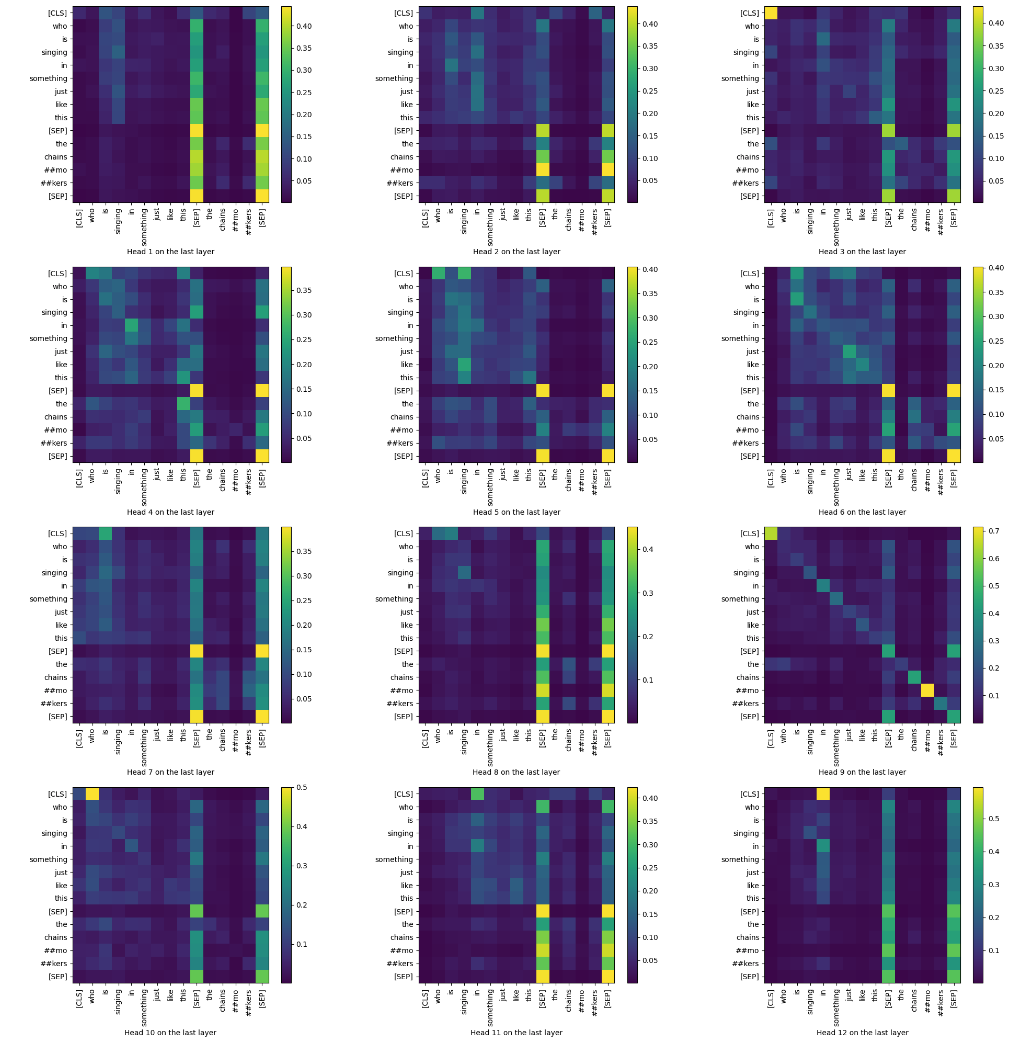}
    \caption{The attention scores of the heads in the last layer of the binary context reranker. There is high attention on [SEP] token across the heads. However, this high attention on the [SEP] token does not provide significant insights into the given context.}
    \label{fig:aslast}
\end{figure*}



\end{document}